\documentclass[sigconf]{acmart}
\copyrightyear{2023}
\acmYear{2023}
\setcopyright{rightsretained}
\acmConference[ASPDAC '23]{28th Asia and South Pacific Design
Automation Conference}{January 16--19, 2023}{Tokyo, Japan}
\acmBooktitle{28th Asia and South Pacific Design Automation Conference
(ASPDAC '23), January 16--19, 2023, Tokyo,
Japan}\acmDOI{10.1145/3566097.3568350}
\acmISBN{978-1-4503-9783-4/23/01}

\usepackage{amsmath}
\usepackage{mathtools}

\usepackage[text-rm]{siunitx}
\usepackage{booktabs}
\usepackage{multirow}
\usepackage{arydshln}
\usepackage{tabularx}
\usepackage{xspace}
\usepackage{stmaryrd}
\usepackage{wrapfig}

\usepackage[scaled]{helvet}

\newcommand{\ignore}[1]{}

\newcommand{\jk}[1]{{\color{black}#1}}
\newcommand{\jktwo}[1]{{\color{black}#1}}
\newcommand{\jkthree}[1]{{\color{black}#1}}
\newcommand{\jkfour}[1]{{\color{black}#1}}

\newcommand{\agy}[1]{\textcolor{blue}{#1}}
\newcommand{\agycomment}[1]{\textcolor{blue}{\textbf{[@agy:} #1\textbf{]}}}

\newcommand{\atb}[1]{\textcolor{red}{#1}}
\newcommand{\atbcomment}[1]{\textcolor{red}{\textbf{[@atb:} #1\textbf{]}}}

\newcommand{\om}[1]{\textcolor{purple}{#1}}
\newcommand{\omcomment}[1]{\textcolor{purple}{\textbf{[@om:} #1\textbf{]}}}

\newcommand{\exploitingRowHammerAllCitations}[0]{\cite{rowhammer-js, glitch-vu, fournaris2017exploiting, poddebniak2018attacking, tatar2018throwhammer, carre2018openssl, barenghi2018software, zhang2018triggering, bhattacharya2018advanced, google-project-zero, google-rh-blackhat, kim2014flipping, rowhammergithub, seaborn2015exploiting, van2016drammer, gruss2016rowhammer, razavi2016flip, pessl2016drama, xiao2016one, bosman2016dedup, bhattacharya2016curious, burleson2016invited, qiao2016new, brasser2017can, jang2017sgx, aga2017good, mutlu2017rowhammer, tatar2018defeating, gruss2018another, lipp2018nethammer, van2018guardion, frigo2018grand, cojocar2019eccploit,  ji2019pinpoint, mutlu2019rowhammer, hong2019terminal, kwong2020rambleed, frigo2020trrespass, cojocar2020rowhammer, weissman2020jackhammer, zhang2020pthammer, yao2020deephammer, deridder2021smash, hassan2021utrr, jattke2022blacksmith, tol2022toward, kogler2022half, orosa2022spyhammer, zhang2022implicit, liu2022generating, cohen2022hammerscope, zheng2022trojvit, fahr2022frodo, tobah2022spechammer, rakin2022deepsteal}}

\newcommand{\exploitingRowHammerNewCitations}[0]{\cite{kwong2020rambleed, frigo2020trrespass, cojocar2020rowhammer, weissman2020jackhammer, zhang2020pthammer, yao2020deephammer, deridder2021smash, hassan2021utrr, jattke2022blacksmith, tol2022toward, kogler2022half, orosa2022spyhammer, zhang2022implicit, liu2022generating, cohen2022hammerscope, zheng2022trojvit, fahr2022frodo, tobah2022spechammer, rakin2022deepsteal}}

\newcommand{\understandingRowHammerNewCitations}[0]{\cite{kim2020revisiting, walker2021ondramrowhammer, orosa2021deeper, orosa2022spyhammer, cohen2022hammerscope, yaglikci2022understanding, khan2018analysis, agarwal2018rowhammer, ni2018write, genssler2022reliability}}

\newcommand{\mitigatingRowHammerNewCitations}[0]{\cite{park2020graphene, yaglikci2021security, devaux2021method, yaglikci2021blockhammer, frigo2020trrespass, kang2020cattwo, hassan2021utrr, qureshi2022hydra, saileshwar2022randomized, kim2022mithril, lee2021cryoguard, marazzi2022protrr, zhang2022softtrr, joardar2022learning, juffinger2023csi, yaglikci2022hira, saxena2022aqua, enomoto2022efficient, manzhosov2022revisiting, ajorpaz2022evax, naseredini2022alarm, joardar2022machine, hassan2022case, zhang2020leveraging, loughlin2021stop, fakhrzadehgan2022safeguard, saroiu2022price, loughlin2022moesiprime, park2022rowhammer, han2021surround, zhou2022ltpim, saroiu2022configure}}

\renewcommand\footnotetextcopyrightpermission[1]{}

\usepackage{algorithm}
\usepackage{algorithmicx}
\floatname{algorithm}{Template}

\usepackage{caption}
\usepackage{subcaption}
\usepackage[normalem]{ulem}

\usepackage{setspace}
\usepackage{fancyhdr}

\begin{document}
\title{Fundamentally Understanding and Solving RowHammer}

\author{Onur Mutlu}
\email{onur.mutlu@safari.ethz.ch}
\orcid{0000-0002-0075-2312}
\affiliation{%
  \institution{ETH Zürich}
  \streetaddress{Gloriastrasse 35}
  \city{Zürich}
  \country{Switzerland}
  \postcode{8092}
}
\author{Ataberk Olgun}
\email{ataberk.olgun@safari.ethz.ch}
\orcid{0000-0001-5333-5726}
\affiliation{%
  \institution{ETH Zürich}
  \streetaddress{Gloriastrasse 35}
  \city{Zürich}
  \country{Switzerland}
  \postcode{8092}
}
\author{A. Giray Ya\u{g}l{\i}kc{\i}}
\email{giray.yaglikci@safari.ethz.ch}
\orcid{0000-0002-9333-6077}
\affiliation{%
  \institution{ETH Zürich}
  \streetaddress{Gloriastrasse 35}
  \city{Zürich}
  \country{Switzerland}
  \postcode{8092}
}

\renewcommand{\shortauthors}{Mutlu et al.}

\begin{abstract}

We provide an overview of recent developments and future directions in the RowHammer vulnerability that plagues modern DRAM (Dynamic Random Memory Access) chips, which are used in almost all computing systems as main memory. 

RowHammer is the phenomenon in which repeatedly accessing a row in a real DRAM chip causes bitflips (i.e., data corruption) in physically nearby rows. This phenomenon leads to a serious and widespread system security vulnerability, as many works since the original RowHammer paper in 2014 have shown. Recent analysis of the RowHammer phenomenon reveals that the problem is getting much worse as DRAM technology scaling continues: newer DRAM chips are fundamentally more vulnerable to RowHammer at the device and circuit levels. Deeper analysis of RowHammer shows that there are many dimensions to the problem as the vulnerability is sensitive to many variables, including environmental conditions (temperature \& voltage), process variation, stored data patterns, as well as memory access patterns and memory control policies. As such, it has proven difficult to devise fully-secure and very efficient (i.e., low-overhead in performance, energy, area) protection mechanisms against RowHammer and attempts made by DRAM manufacturers have been shown to lack security guarantees. 

After reviewing various recent developments in exploiting, understanding, and mitigating RowHammer, we discuss future directions that we believe are critical for solving the RowHammer problem. We argue for two major directions to amplify research and development efforts in: 1) building a much deeper understanding of the problem and its many dimensions, in both cutting-edge DRAM chips and computing systems deployed in the field, and 2) the design and development of extremely efficient and fully-secure solutions via system-memory cooperation.

\end{abstract}

\begin{CCSXML}
<ccs2012>
   <concept>
       <concept_id>10010583.10010600.10010607.10010608</concept_id>
       <concept_desc>Hardware~Dynamic memory</concept_desc>
       <concept_significance>500</concept_significance>
       </concept>
   <concept>
       <concept_id>10002978.10003006</concept_id>
       <concept_desc>Security and privacy~Systems security</concept_desc>
       <concept_significance>500</concept_significance>
       </concept>
   <concept>
       <concept_id>10010583.10010750.10010762</concept_id>
       <concept_desc>Hardware~Hardware reliability</concept_desc>
       <concept_significance>500</concept_significance>
       </concept>
 </ccs2012>
\end{CCSXML}

\ccsdesc[500]{Hardware~Dynamic memory}
\ccsdesc[500]{Security and privacy~Systems security}
\ccsdesc[500]{Hardware~Hardware reliability}

\keywords{DRAM, Security, Vulnerability, Technology Scaling, Reliability, Safety, Errors, Memory Systems, Fault Attacks, RowHammer}

\maketitle

% \fancypagestyle{plain}{%
% \fancyhf{} % clear all header and footer fields
% \fancyfoot[C]{\fontsize{10pt}{10pt}\selectfont\thepage} % except the center
% \renewcommand{\headrulewidth}{0pt}
% \renewcommand{\footrulewidth}{0pt}}

% \pagestyle{plain}
% \thispagestyle{plain}
\settopmatter{printfolios=true}

% \vspace{-1em}
\section{Introduction}

%%% CHECK-AG: Check below and fix references (including sections, citations, etc.)

DRAM is the dominant technology used for main memory in almost all computing systems due to its low latency and low cost per bit. Modern DRAM chips suffer from a vulnerability commonly known as RowHammer~\cite{kim2014flipping,mutlu2017rowhammer,mutlu2019rowhammer, mutlu2019retrospective}. RowHammer is caused by repeatedly accessing (i.e., hammering) one or more (aggressor) memory rows. Hammering a row creates electromagnetic interference between the aggressor row and its physically-neighboring (victim) rows. Due to this interference, cells in victim rows lose the ability to correctly retain their data, which leads to data corruption (i.e., bitflips). These bitflips are repeatable: if hammering an aggressor row causes a particular cell to experience a bitflip, doing so again will lead to the same bitflip with high probability~\cite{kim2014flipping}. 

Unfortunately, DRAM becomes increasingly more susceptible to RowHammer bitflips as its storage density increases (i.e., DRAM cell size and cell-to-cell spacing reduce). Our recent work~\cite{kim2020revisiting} across 1580 real DRAM chips of six different types shows that, in the last decade, the minimum number of aggressor row activations needed to cause a RowHammer bitflip (i.e., the {\em RowHammer threshold}) has reduced by more than 10x and the number of bitflips caused by the same number of row activations has increased by 500x. A recent work~\cite{loughlin2022moesiprime} shows that commodity workloads on state-of-the-art servers might already activate individual DRAM rows at rates exceeding the RowHammer threshold.

%the minimum number of aggressor row activations needed to cause a RowHammer bitflip.
\ignore{
\agycomment{should we add RRS, AQUA, Hydra, and BlockHammer here for consistency? It would require to rewrite the sentence and make it weaker as these works do not analyze benign workload access patterns on state-of-the-art servers.}
}

On the one hand, such RowHammer bitflips can lead to system reliability and safety problems, when caused by non-malicious applications. On the other hand, malicious applications can be written to induce RowHammer bitflips in a targeted manner, so as to specifically degrade system security, privacy, safety and availability. For example, by carefully selecting rows to hammer, an attacker can induce bitflips in sensitive data stored in DRAM. Many prior works, some of which are reviewed in~\cite{mutlu2019retrospective}, show that RowHammer can be exploited in many ways to compromise system security (integrity, confidentiality, availability) in real systems, since it breaks {\em memory isolation} on top of which modern system security principles are built. As such, RowHammer greatly threatens many aspects of computing system robustness (which include system reliability, safety, security, privacy, and availability) in a widespread and profound manner due to the prevalent usage of DRAM in modern computing systems. 

%%% and thus has widespread and profound real implications

RowHammer-like disturbance issues have also been shown to be present in emerging NVM (non-volatile memory) technologies~\cite{khan2018analysis, agarwal2018rowhammer, li2014write, ni2018write, genssler2022reliability}. For these technologies to be viable, dependable, and secure, anticipation and solution of such issues are critical. 

In this paper, we provide an overview of the state-of-the-art research and development focusing on the RowHammer vulnerability. To this end, we first provide a very brief review of RowHammer research until circa 2020 (Section~\ref{sec:rowhammer-problem}), building on a recent overview paper~\cite{mutlu2019retrospective} that in more detail covers much of the RowHammer developments until its publication. Then, we describe two major developments in 2020 (Section~\ref{sec:recent1}), TRRespass~\cite{frigo2020trrespass} and Revisiting RowHammer~\cite{kim2020revisiting}, which  experimentally show that RowHammer is an open and increasingly worsening problem, which motivated a large body of follow-on work in both industry and academia. Afterwards, we provide a broad overview of other recent developments in RowHammer (Section~\ref{sec:recent2}). We consider three major types of RowHammer developments: 1) works directed toward exploiting RowHammer, 2) works that aim to understand and model RowHammer, and 3) works that propose techniques to mitigate/solve the RowHammer problem. Finally, we discuss future directions in RowHammer research (Section~\ref{sec:future}). We argue for two major broad directions to focus more in future research and development efforts: 1) building a much deeper understanding of the problem and its many dimensions/sensitivities, in both cutting-edge DRAM chips and computing systems deployed in the field, and 2) the design and development of extremely efficient and fully-secure architectural solutions, which we believe can be achieved via much better system-memory cooperation (as discussed in~\cite{mutlu2013memory, mutlu2015research}).

\ignore{
We argue for two major directions to amplify research and development efforts in: 1) building a much deeper understanding of the problem and its many dimensions, in both cutting-edge DRAM chips and computing systems deployed in the field, and 2) the design and development of extremely efficient and fully-secure solutions via system-memory cooperation. 
}

\ignore{ ONUR: Removed these citations, nnecessary at this point and not clear they make sense.
~\cite{loughlin2022moesiprime,saxena2022aqua,saileshwar2022randomized,qureshi2022hydra} \atbcomment{giray can you add here?}\agycomment{I'm not sure what to add... MOESI-Prime maybe?} \agycomment{AQUA, RRS, and Hydra have some similar analysis but they don't necessarily suggest this sentence.}\omcomment{Is there anything more to add here that talks about non-malicious applications? I would } 
}

\ignore{
Memory is a key component of all modern computing systems, often
determining the overall performance, energy efficiency, and
reliability characteristics of the entire system. The push for
increasing the density of modern memory technologies via technology
scaling, which has resulted in higher capacity (i.e., density) memory
and storage at lower cost, has enabled large leaps in the performance
of modern computers~\cite{mutlu-imw13}. This positive trend is clearly
visible in especially the dominant main memory and solid-state storage
technologies of today, i.e., DRAM~\cite{tldram, lee2017design,
  kevinchang-sigmetrics16, kim2018solar, chang_Understanding2017} and
NAND flash memory~\cite{cai-date12, cai-iccd13, cai2017error},
respectively.  Unfortunately, the same push has also greatly decreased
the reliability of modern memory technologies, due to the increasingly
smaller memory cell size and increasingly smaller amount of charge
that is maintainable in the cell, which makes the memory cell much
more vulnerable to various failure mechanisms and noise and
interference sources, both in DRAM~\cite{dram-isca2013,
  rowhammer-isca2014, samira-sigmetrics14, kang-memforum2014,
  avatar-dsn15, khan-dsn16, memcon-cal16, memcon-micro17, onur-date17,
  patel2017reach} and NAND flash nemory~\cite{cai2017error,
  cai-date12, cai-date13, cai-iccd13, cai-hpca15, cai-hpca17,
  cai-iccd12, cai-itj2013, cai-dsn15, cai-sigmetrics14, yixin-jsac16,
  onur-date17, luo2018improving, luo2018heatwatch, cai2017errors,
  warm-msst15}.

As memory scales down to smaller technology nodes, new failure
mechanisms emerge that threaten its correct operation. If such failure
mechanisms are not anticipated and corrected, they can not only
degrade system reliability and availability but also, perhaps even
more importantly, open up new security vulnerabilities: a malicious
attacker can exploit the exposed failure mechanism to take over the
entire system. As such, new failure mechanisms in memory can become
practical and significant threats to system security.

In this article, we provide a retrospective of one such example
failure mechanism in DRAM, which was initially introduced by Kim et
al. at the ISCA 2014 conference~\cite{rowhammer-isca2014}. We provide
a description of the RowHammer problem and its implications by
summarizing our ISCA 2014 paper~\cite{rowhammer-isca2014}, describe
solutions proposed by our original work~\cite{rowhammer-isca2014},
comprehensively examine the many works that build on our original work
in various ways, e.g., by developing new security attacks, proposing
solutions, and analyzing RowHammer. What comes next in this section
provides a roadmap of the entire article.

In our ISCA 2014 paper~\cite{rowhammer-isca2014}, we introduce the
RowHammer problem in DRAM, which is a prime (and perhaps the first)
example of how a circuit-level failure mechanism can cause a
practical and widespread system security vulnerability.  RowHammer, as
it is now popularly referred to, is the phenomenon that repeatedly
accessing a row in a modern DRAM chip causes bit flips in
physically-adjacent rows at consistently predictable bit locations. It
is caused by a hardware failure mechanism called DRAM disturbance
errors, which is a manifestation of circuit-level cell-to-cell
interference in a scaled memory technology. We describe the RowHammer
problem and its root causes in Section~\ref{sec:rowhammer-problem}.

Inspired by our ISCA 2014 paper's fundamental findings, researchers
from Google Project Zero demonstrated in 2015 that this hardware
failure mechanism can be effectively exploited by user-level programs
to gain kernel privileges on real
systems~\cite{google-project-zero,google-rh-blackhat}. Tens of other
works since then demonstrated other practical attacks exploiting
RowHammer, e.g., ~\cite{cloudflops, dedup-est-machina, anotherflip,
  qiao2016new, bhattacharya2016curious, jang2017sgx, aga2017good,
  pessl2016drama, rowhammer-js, flip-feng-shui, drammer,
  glitch-vu,fournaris2017exploiting, poddebniak2018attacking, nethammer, throwhammer,
  tatar2018defeating, carre2018openssl,
  barenghi2018software, zhang2018triggering,
  bhattacharya2018advanced, cojocar19exploiting}. These include remote takeover of a server vulnerable to
RowHammer, takeover of a victim virtual machine by another virtual
machine running on the same system, takeover of a mobile device by a
malicious user-level application that requires no permissions,
takeover of a mobile system quickly by triggering RowHammer using a
mobile GPU, and takeover of a remote system by triggering RowHammer on
it through the Remote Direct Memory Access (RDMA)
protocol~\cite{rdma-consortium}. We describe the works that build on RowHammer
to develop new security attacks in Section~\ref{sec:related-exploits}.
%%% JEREMIE-DONE: Please add reference to the RDMA protocol above. 

Our ISCA 2014 paper rigorously and experimentally analyzes the
RowHammer problem and examines seven different solutions, multiple of
which are already employed in practice to prevent the security
vulnerabilities (e.g., increasing the memory refresh rate).  We
propose a low-cost solution, Probabilistic Adjacent Row Activation,
which provides a strong and configurable reliability and security
guarantee; a solution whose variants are being adopted by DRAM
manufacturers and memory controller designers~\jk{\cite{x210-rh-ss}}. We describe this
solution and the six other solutions of our original paper in
Section~\ref{sec:rowhammer-solutions}. Many other works build on our
original paper to propose and evaluate other solutions to RowHammer,
and we discuss them comprehensively in
Section~\ref{sec:related-defenses}.

Our ISCA 2014 paper leads to a new mindset that has enabled a renewed
interest in hardware security research: general-purpose hardware is fallible, in a very widespread manner, and this causes real security problems.  We believe the RowHammer problem will
become worse over time since DRAM cells are getting closer to each
other with technology scaling. Other similar vulnerabilities may also
be lurking in DRAM and other types of memories, e.g., NAND flash
memory or Phase Change Memory, that can potentially threaten the
foundations of secure systems, as the memory technologies scale to
higher densities. Our ISCA 2014 paper advocates a principled
system-memory co-design approach to memory reliability and security
research that can enable us to better anticipate and prevent such
vulnerabilities. We describe promising ongoing and future research
directions related to RowHammer (Section~\ref{sec:future}), including
the examination of other potential vulnerabilities in memory (in
Section~\ref{sec:future-other-problems}) and the use of a principled
approach to make memory more reliable and more secure (in
Section~\ref{sec:future-prevention}).

%%% JEREMIE-DONE: Could you fix the references to sections above?

%%% ONUR-DONE: Ensure consistency among the references and sections.
}
\section{A Brief Overview of RowHammer Until 2020}
\label{sec:rowhammer-problem}

%% TODO-OM: Brief overview of original work and a summary of many works, with heavy references to the TCAD retrospective for further information. 

Since its first public introduction and scientific analysis in 2014~\cite{kim2014flipping}, RowHammer has led to significant follow-on work in both academia and industry, spanning multiple different communities, including computer architecture, security, dependability, circuits, devices, and systems. We briefly review works that appeared within the timeframe 2014-2019. A more comprehensive overview of such works, as well as others in popular technical media, can be found in~\cite{mutlu2019retrospective}, and we refer the reader to that overview for more detail.

The ISCA 2014 work that introduced RowHammer~\cite{kim2014flipping} demonstrated that more than 80\% of the real commodity DDR3 DRAM modules tested, from all three major DRAM manufacturers, are vulnerable to RowHammer. That is, real bitflips are possible to induce using real user-level programs~\cite{rowhammergithub} on commonly-used CPU-based systems. The work argued that RowHammer is a DRAM technology scaling problem and since device- and circuit-level solutions are difficult and costly, RowHammer should be solved via system-memory cooperation~\cite{kim2014flipping,mutlu2013memory}. This work also suggested that one can exploit RowHammer bitflips to construct various types of {\em disturbance attacks} that inject errors into other programs, crash the system, or hijack control of the system. 

%%% CHECK-AG: Please check the references below and make sure they match the exploit works covered in Section III-A of~\cite{mutlu2019rowhammer}. The below references seem incomplete compared to what we cite in that section. I want to exactly match that section + add any missing works in that section.

Many future works building on Kim et al.~\cite{kim2014flipping} did exactly that, i.e., they developed various types of attacks that exploit RowHammer on real computing systems. These works include techniques that compromise system integrity as well as confidentiality on various types of systems, including mobile and server systems~\cite{seaborn2015exploiting, van2016drammer, gruss2016rowhammer, razavi2016flip, pessl2016drama, xiao2016one, bosman2016dedup, bhattacharya2016curious, burleson2016invited, qiao2016new, brasser2017can, jang2017sgx, aga2017good, mutlu2017rowhammer, tatar2018defeating, gruss2018another, lipp2018nethammer, van2018guardion, frigo2018grand, cojocar2019eccploit,  ji2019pinpoint, mutlu2019rowhammer, hong2019terminal, kwong2020rambleed,  fournaris2017exploiting, poddebniak2018attacking, tatar2018throwhammer, carre2018openssl, barenghi2018software, zhang2018triggering, bhattacharya2018advanced, google-project-zero}. A more detailed overview of these works can be found in Section III-A of~\cite{mutlu2019retrospective}. 

% these are from seciii-a in retrospective 
%  cloudflops--> xiao2016one, dedup-est-machina-->bosman2016dedup, anotherflip-->gruss2018another,
%   qiao2016new-->qiao2016new, bhattacharya2016curious-->bhattacharya2016curious, jang2017sgx->jang2017sgx, aga2017good-->aga2017good,
%   pessl2016drama-->pessl2016drama, rowhammer-js --> rowhammer-js, flip-feng-shui-->razavi2016flip, drammer-->van2016drammer,
%   glitch-vu-->glitch-vu,fournaris2017exploiting--> fournaris2017exploiting, poddebniak2018attacking-->poddebniak2018attacking, nethammer-->lipp2018nethammer, throwhammer-->tatar2018throwhammer,
%   tatar2018defeating-->tatar2018defeating, carre2018openssl-->carre2018openssl,
%   barenghi2018software-->barenghi2018software, zhang2018triggering-->zhang2018triggering,
%   bhattacharya2018advanced-->bhattacharya2018advanced, cojocar19exploiting-->cojocar2019eccploit, google-project-zero-->google-rh-blackhat,
%   google-rh-blackhat

%%%% CHECK-AG: Add references to device/circuit-level "understanding" works here from appropriate sections of our retrospective paper. I will write the text afterwards.

Multiple works at the device and circuit levels aimed to develop a low-level understanding of the causes and effects of RowHammer via low-level modeling and simulation of devices and circuits, and in limited cases via experiments on DDR3 DRAM chips. These works include~\cite{park2016experiments,park2016statistical,ryu2017overcoming,yun2018study,yang2019trap}. A more recent work in 2021~\cite{walker2021ondramrowhammer} complements this understanding with some newer device level observations. Similarly, RowHammer-like behavior has been analyzed in NVM (non-volatile memory) devices via device and circuit level simulation studies~\cite{khan2018analysis, agarwal2018rowhammer, li2014write, ni2018write, genssler2022reliability}. We refer the reader to Sections III-C and III-H of~\cite{mutlu2019retrospective} for a more detailed overview of such device/circuit level works that aim to develop a better low-level understanding of the causes and effects of RowHammer. 

\ignore{

\atb{\sout{most}} 

\omcomment{Ataberk \& Giray: can you check and references to circuit and device level works from after 2020? I would like to see what they are so that we can decide where to place these. The above paragraph now mixes works from 2014-2019 with a few later ones...} \atbcomment{We found circuit/device-level works that do TCAD simulations, but they propose defenses and evaluate them using TCAD. These are:2020+~\cite{park2022rowhammer, han2021surround}, before~\cite{gautam2018improvement}. Not sure if these count. We added them to defenses though. Besides these, walker paper~\cite{walker2021ondramrowhammer} seems to be the only one from 2019+}\omcomment{Can you make sure references you found from 2020 and beyond are added to the later section where we discuss "understanding". Please ack after you confirm you did that so that we can remove these comments.} \atbcomment{ACK}
}

%New found device level work (before 2020):~\cite{gautam2018improvement}\atbcomment{More to come}

\ignore{ONUR: I added all of these to the above paragraph
\cite{yang2019trap}: 3D CAD 
\cite{yun2018study}: Effects of irradiating DRAM
\cite{lim2017active}: Irradiating with gamma rays
\cite{ryu2017overcoming}: hydrogen annealing
\cite{park2016experiments} and \cite{park2016statistical}: Experimental DDR3 tests.
Walker paper \cite{walker2021ondramrowhammer}.
NVM RowHammer understanding~\cite{khan2018analysis, agarwal2018rowhammer, li2014write, ni2018write, genssler2022reliability}.
}

%%%% CHECK-AG: Add references to "mitigating" works here from appropriate sections of our retrospective paper. I will write the text afterwards

The original RowHammer paper in ISCA 2014~\cite{kim2014flipping} proposed seven different solution directions to RowHammer, several of which were later implemented in variations in memory controllers and DRAM chips. Building on these, many other academic and industrial works in the 2014-2019 timeframe proposed various solutions to RowHammer in both hardware and software. These works include~\cite{bains-merged, bains14d, bains14c, greenfield14a, AppleRefInc,rh-hp,rh-lenovo,rh-cisco, aweke2016anvil, kim2014architectural, seyedzadeh2017counter, brasser2017can, irazoqui2016mascat, son2017making, gomez2016dram, van2018guardion, lee2018twice, lee2019twice, bu2018srasa, oh2018reliable, konoth2018zebram, izzo2017reliably, google-project-zero, wu2019protecting, bains2016row}. A detailed overview of these solutions can be found in Section III-B of~\cite{mutlu2019retrospective}. The BlockHammer paper in HPCA 2021~\cite{yaglikci2021blockhammer} provides a more up-to-date overview of major solutions proposed until that time, with a detailed analysis of \om{14} solutions across four different desirable properties and a rigorous evaluation of six state-of-the-art solutions, along with an open-source release of their source codes~\cite{blockhammergithub}.  

\ignore{
\omcomment{Please check the above description and add citations. Also make sure no references are repeated or missing.}\atbcomment{done}
\omcomment{ALso: are we missing any patents from that timeframe???}\atbcomment{patents should be complete (assuming revising rowhammer is complete), these are~\cite{bains2015row, bains2015rowrefcmd, bains14d, bains14c, bains14a,
  bains14b, greenfield14a}}\omcomment{OK for now.}
  }

\ignore{ONUR: I added all of these to the above paragraph
Original paper~\cite{kim2014flipping}.
"Academic works" in retrospective (referenced as examples)~\cite{aweke2016anvil, kim2014architectural, seyedzadeh2017counter, seyedzadeh2017counter, brasser2017can, irazoqui2016mascat,
  son2017making, gomez2016dram, van2018guardion, lee2018twice, lee2019twice,
  bu2018srasa, oh2018reliable}.
Immediate software solutions in retrospective~\cite{konoth2018zebram,
  izzo2017reliably, brasser2017can, irazoqui2016mascat,
  google-project-zero, van2018guardion, wu2019protecting, oh2018reliable}.
"Long-term" solutions in retrospective~\cite{gomez2016dram, bu2018srasa, gong2018memory, jones2017holistic, kline2017sustainable, schilling2018pointing, seyedzadeh2017counter, seyedzadeh2018cbt, lee2018twice, lee2019twice, son2017making, danger2018ccfi, bains2015row, bains14d, greenfield14a, wang2019detect, vig2018rapid, kim2019effective}.
Patents in retrospective~\cite{bains2015row, bains14d, bains14c, bains14a, bains14b, greenfield14a}.
Increased refresh rate in retrospective~\cite{AppleRefInc,rh-hp,rh-lenovo,rh-cisco}.
}

%%%% TODO-AG: PLease add any other suggestions you may have, as comments. I will return to this section later

\vspace{-1em}
\subsection{RowHammer Mitigations in Industry}

After the public introduction of RowHammer in 2014, both system and DRAM manufacturers took action to mitigate the problem in the field and in future DRAM chips. As described in the original RowHammer work~\cite{kim2014flipping}, the solutions that can be deployed in the field are limited due to the limited programmability support provided by modern memory controllers. As such, a major solution deployed in the field has been to increase the refresh rate of DRAM, as described in a security release by Apple~\cite{AppleRefInc}. Unfortunately, increasing the refresh rate is not an effective or desirable solution~\cite{kim2014flipping} due to its high performance and energy overheads~\cite{liu2012raidr}.

On the system side, memory controller manufacturers like Intel introduced   mechanisms  (e.g., ~\cite{x210-rh-ss}, inspired by {\em PARA, probabilistic adjacent row activation}~\cite{kim2014flipping}), broadly called {\em pTRR} (pseudo Target Row Refresh)~\cite{frigo2020trrespass, kaczmarski2014thoughts}, to mitigate RowHammer in future systems. Unfortunately, such memory controller based victim row refresh mechanisms are not aware of physical adjacency of aggressor and victim rows in a DRAM chip and, as such, they might not provide complete protection against RowHammer.

On the DRAM side, DRAM manufacturers introduced TRR (target row refresh)~\cite{frigo2020trrespass} mechanisms and claimed that their new DDR4 chips are RowHammer-free~\cite{frigo2020trrespass, hassan2021utrr, lee2014green, micron2014ddr4} with the protection provided by these mechanisms. TRR is an umbrella term used for mechanisms that refresh target rows which are somehow determined to be accessed frequently. Unfortunately, DRAM manufacturers did not (and still do not) describe how their implementations securely prevent RowHammer or reveal how their implementations work. 

As such, circa 2019-2020, it was unclear whether or not RowHammer bitflips were possible in real DDR4 DRAM chips. As we will see next, two major works in 2020 showed that they were.

\section{Major Developments in 2020}
\label{sec:recent1}

%% CHECK-OM: More detail for recent works

We first describe two major works, {\em TRRespass}~\cite{frigo2020trrespass} and {\em Revisiting RowHammer}~\cite{kim2020revisiting}, which clearly demonstrate that RowHammer is an open and worsening problem, state-of-the-art DRAM chips are vulnerable, and proposed mitigations are not scalable/effective into the future.

{\em TRRespass}~\cite{frigo2020trrespass} is the first work to show that TRR-protected DDR4 DRAM chips that are advertised as Rowhammer-free are actually vulnerable to RowHammer in the field. This work partially reverse engineers the TRR and pTRR mechanisms employed in modern DRAM chips and memory controllers. To overcome such protection mechanisms, TRRespass introduces the {\em many-sided RowHammer attack}, whose key idea is to hammer {\em many} (i.e., more than two) rows to bypass TRR mitigations, e.g., by overflowing proprietary TRR tables that detect aggressor rows. Using this many-sided RowHammer attack, the work demonstrated bitflips in real DDR4 DRAM chips as well as LPDDR4(X) DRAM chips and showed that RowHammer attacks are possible on systems that employ such chips. As such, it was clear that the solutions implemented in industry were not secure. TRRespass also argued that {\em security by obscurity}, as employed by DRAM manufacturers, is not a good solution approach. Later follow-on work, called Uncovering TRR (U-TRR)~\cite{hassan2021utrr}, showed in 2021 that one can almost completely reverse engineer the entire TRR mechanism employed in any DRAM chip, by using an FPGA-based DRAM testing infrastructure (i.e., SoftMC~\cite{hassan2017softmc, softmcgithub} and DRAM Bender~\cite{olgun2022drambender}) and a methodology that uses retention errors as side channels to discover when the DRAM-internal TRR mechanism refreshes a victim row. U-TRR demonstrates that, by doing so, one can craft specialized hammering/access patterns that essentially induce large numbers of bitflips on any examined chip.

\begin{figure}[h]
    \centering
    \includegraphics[width=\linewidth]{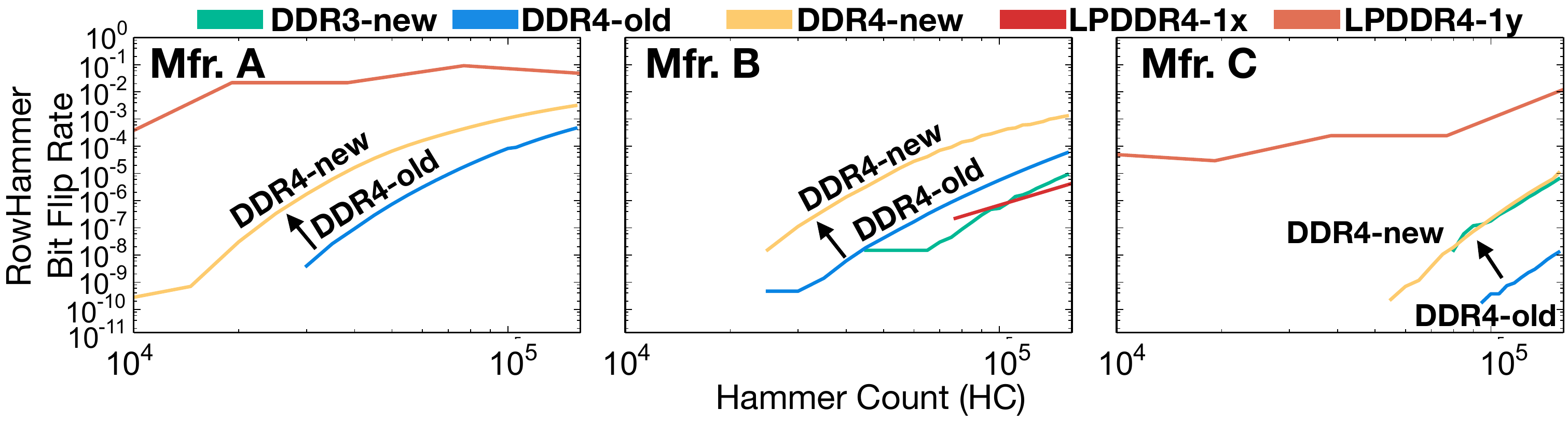}
    \vspace{-2em}
    \caption{Hammer count (HC) vs. RowHammer bitflip rate across 1580 DRAM chips from five different type-node configurations. Reproduced from~\cite{kim2020revisiting}.}
    \vspace{-1em}
    \label{fig:hcfirst-ber}
\end{figure}

{\em Revisiting RowHammer}~\cite{kim2020revisiting} takes a device/circuit-level approach to understand the scaling properties of RowHammer by measuring the fundamental vulnerability (i.e., with TRR mechanisms turned off) of three different types of DRAM chips across at least two different generations. This work tested 1580 DRAM chips and experimentally demonstrated that RowHammer is {\em indisputably getting worse} in newer generation DRAM chips: when hammered, newer DRAM chips experience the first bitflip much earlier (e.g., some after only 4800 double-sided hammers) and they experience much higher numbers of bitflips than older DRAM chips (as demonstrated in Figure ~\ref{fig:hcfirst-ber}\footnote{Figure~\ref{fig:hcfirst-ber} (reproduced from Revisiting RowHammer~\cite{kim2020revisiting}) illustrates a quantitative summary of both the reduction in double-sided hammer count (x-axis) and the increase in RowHammer bitflip rate (y-axis) with newer DRAM generations (e.g., DDR4-old in blue to DDR4-new in yellow) across three major DRAM manufacturers (A, B, C).}). In other words, the number of activations to induce a RowHammer bitflip (i.e., the {\em RowHammer threshold}) reduced from 139K single-sided in 2014~\cite{kim2014flipping} to 4.8K double-sided in 2020~\cite{kim2020revisiting}. Revisiting RowHammer also showed that if the scaling trend continues as such, all known solutions at the time would either not work in future DRAM chips that are even more vulnerable (e.g., with a RowHammer threshold of 256 or 128) or have prohibitively large performance overheads. To our knowledge, this work is the largest scaling study of RowHammer to date, covering many different types and generations of DRAM chips and its results demonstrate the criticality and difficulty of the RowHammer problem as technology node size shrinks in DRAM manufacturing.

Alarmed by the key experimental results and takeaways demonstrated by these two works~\cite{frigo2020trrespass, kim2020revisiting}, industry took action to gather forces to more seriously mitigate the RowHammer problem. To this end, a major RowHammer task group was (re)organized in JEDEC (Joint Electron Device Engineering Council~\cite{jedecwebsite}). This task group produced whitepapers describing how to proceed to solve RowHammer~\cite{jedec2021system, jedec2021nearterm}, which argue for limited system-DRAM cooperation. The mission of this task group continues.

JEDEC and DRAM manufacturers also added a new RFM (Refresh Management) feature to the DDR5 standard~\cite{jedec2020ddr5} to aid the in-DRAM mitigation mechanisms. The key idea of RFM is to provide in-DRAM TRR mechanisms with extra time (as needed) to securely refresh all potential victim rows. RFM requires the memory controller to count the number of activations at DRAM bank granularity and issue an RFM refresh command when the activation count reaches a threshold value (based on the RowHammer threshold and the strength of the in-DRAM TRR mechanism of a DRAM chip), such that a RowHammer defense mechanism implemented inside the DRAM chip can refresh the victim rows. However, because the memory controller tracks row activation counts at DRAM bank granularity, the activation count can frequently reach the threshold even when there is no RowHammer attack in the system. As a result, the memory controller can issue many unnecessary RFM commands, each of which makes a DRAM bank unavailable for hundreds of nanoseconds, causing performance overhead.

\section{Recent RowHammer Developments}
\label{sec:recent2}

TRRespass~\cite{frigo2020trrespass} and Revisiting RowHammer~\cite{kim2020revisiting}, by demonstrating the importance and openness of the problem, catalyzed the ongoing RowHammer research and development efforts. We briefly review such efforts from 2019-2020 until now.

\subsection{Exploiting RowHammer}

Many works since 2019-2020 developed new and different ways to exploit the RowHammer vulnerability (e.g.,~\exploitingRowHammerNewCitations{}). Among these, we briefly describe RAMBleed~\cite{kwong2020rambleed} and RowHammer-driven fault attacks on neural networks~\cite{yao2020deephammer, hong2019terminal, tol2022toward, liu2022generating, rakin2022deepsteal} since they demonstrate different uses of RowHammer from taking over a system. We also briefly describe new attacks that build on many-sided hammering, including SMASH~\cite{deridder2021smash}, Blacksmith~\cite{jattke2022blacksmith}, and Half-Double~\cite{kogler2022half}, as they provide new insights into how vulnerable existing DDR4 DRAM chips are to RowHammer.

RAMBleed~\cite{kwong2020rambleed} shows that RowHammer bitflips can be used to break confidentiality in an existing DRAM-based system: RowHammer bitflips can be used as a side channel to determine the data values stored in a location a user-level program does not have read access to. This work demonstrates an attack against OpenSSH~\cite{openssh} where RAMBleed is used to leak an encryption key.

\sloppypar
Several works~\cite{yao2020deephammer, hong2019terminal, tol2022toward, liu2022generating} demonstrate that targeted RowHammer bitflips can be used to greatly degrade the inference accuracy of a neural network. These works highlight potential safety and security issues that may be caused by RowHammer in systems that rely on neural networks for decision making (e.g., autonomous vehicles). A recent work in 2022~\cite{rakin2022deepsteal} demonstrates that RowHammer bitflips can be used as a side channel to recover the weights stored in a neural network, which breaks confidentiality and violates privacy.

Some works that build on TRRespass demonstrate new RowHammer attacks that are relatively easy to perform on DDR4 DRAM chips. SMASH (Synchronized MAny-Sided Hammering)~\cite{deridder2021smash} successfully triggers RowHammer bitflips from JavaScript code by exploiting many-sided hammering and synchronizing access patterns with DRAM refresh operations to bypass TRR mitigations. This work demonstrates an end-to-end JavaScript exploit to fully compromise the Firefox web browser in 15 minutes. Blacksmith~\cite{jattke2022blacksmith} uses automated fuzzing in the frequency domain to discover non-uniform access patterns that can bypass TRR mechanisms more effectively than uniform many-sided hammering. By doing so, it generates access patterns that hammer aggressor rows with different phases, frequencies, and amplitudes, finding complex patterns that trigger Rowhammer bitflips on all 40 tested DDR4 modules.

Building on the concept of many-sided RowHammer attacks, Google introduced the Half Double hammering pattern in 2021~\cite{kogler2022half, qazi2021introducing, qazi2021halfdoublereport}. Half Double shows that hammering a "far" neighbor row that is physically one row away from the victim row many times and then hammering the immediately-adjacent "near" neighbor row of the victim row a much smaller number of times leads to bitflips in some DDR4 DRAM chips. This is concerning because, when TRR-based RowHammer mitigations are employed, hammers performed on the {\em far} neighbor row can lead to refreshes performed by TRR on the {\em near} neighbor row, which in turn can lead to bitflips in the victim row. This work highlights the intricacies in bitflip mechanisms in modern DRAM chips and alerts that defenses should be carefully designed to work in the presence of such intricacies.

Many other works from 2020-2022 (e.g.,~\cite{cojocar2020rowhammer, weissman2020jackhammer, zhang2020pthammer, orosa2022spyhammer, zhang2022implicit, cohen2022hammerscope, zheng2022trojvit, fahr2022frodo, tobah2022spechammer}) demonstrate various other RowHammer exploits, advancing our understanding of how RowHammer bitflips can be used to degrade system security. We refer the reader to the individual papers for more detail. We believe that it is critical to push the boundaries of system security research by understanding the different ways in which RowHammer bitflips can cause security issues and making RowHammer exploits more powerful, especially in the presence of stronger mitigation mechanisms.

\ignore{
%%% CHECK-AG: Please add references to as many exploit works as possible, from 2020 to 2023. PLease cite them here so that I can write about them. No need to write anything but if there is something special you want to say, please say in comment or text. My recent talks in the Seminar provides a wealth of references and there are more.
\agycomment{I updated the exploitingRowHammer macro below with the citations from TCAD paper.}

Exploiting RowHammer all works~\exploitingRowHammerAllCitations{}.

Exploiting RowHammer post-2020 works~\exploitingRowHammerNewCitations{}.
}

\vspace{-1em}
\subsection{Understanding RowHammer}

Recent works since 2020 (e.g.,~\understandingRowHammerNewCitations{}) study RowHammer from different aspects to develop a better understanding. Among these, we briefly describe two major works: {\em A Deeper Look into RowHammer}~\cite{orosa2021deeper} and {\em RowHammer under Reduced Wordline Voltage}~\cite{yaglikci2022understanding} because these two works analyze new aspects of RowHammer by rigorously testing real DRAM chips.

{\em A Deeper Look into RowHammer}~\cite{orosa2021deeper} presents an experimental characterization using 248 DDR4 and 24 DDR3 modern DRAM chips from four major DRAM manufacturers to reveal how the RowHammer vulnerability is affected by three fundamental properties: 1) DRAM chip temperature, 2) aggressor row active time, and 3) victim DRAM cell’s physical location. The results clearly indicate that a RowHammer bitflip is more likely to occur 1)~in a bounded range of temperature, 2)~if the aggressor row is active for longer time, and 3)~in certain physical regions of the DRAM module under attack. This work also shows how its findings can be used to improve both RowHammer attacks and RowHammer defenses, highlighting the importance of such deeper understanding from the perspective of both attackers and defenders.

{\em RowHammer under Reduced Wordline Voltage}~\cite{yaglikci2022understanding} presents an experimental characterization using 272 real DDR4 DRAM chips from three major manufacturers to demonstrate how reducing the wordline voltage affects both RowHammer vulnerability and DRAM operation. The authors show that reducing wordline voltage provides a significant reduction in the number of RowHammer bitflips and increase in the minimum number of aggressor row activations needed to cause a RowHammer bitflip, without significantly affecting reliable DRAM operation. This work highlights how deeply understanding low-level effects on RowHammer can aid in developing more RowHammer-resilient DRAM systems.

Several other works from 2020-2022 (e.g.,~\cite{walker2021ondramrowhammer, orosa2022spyhammer, cohen2022hammerscope, park2022rowhammer, han2021surround}) present various other simulation and real experiment-based analyses to gain insights into RowHammer and its effects, furthering our fundamental understanding of the RowHammer phenomenon. We refer the reader to the individual papers for more detail. We believe that it is critical to continue to develop a broader and deeper understanding of RowHammer to improve both RowHammer attacks and defenses, on the path to designing systems that can  effectively and efficiently guard against RowHammer.
% \agycomment{I updated the understandingRowHammer macro below with the citations from TCAD paper.}
% %%% CHECK-AG: Ditto here, but for "understanding".

% Understanding RowHammer all works~\understandingRowHammerAllCitations{}.

% Understanding disturbance in NVM works~\understandingRowHammerNVMCitations{}.

% Understanding RowHammer post-2020 works~\understandingRowHammerNewCitations{}.
% \vspace{-1em}
\subsection{Mitigating RowHammer}
\label{sec:mitigating_rowhammer}

{\em Revisiting RowHammer} from ISCA 2020~\cite{kim2020revisiting} clearly demonstrated that, as RowHammer continues to worsen in real DRAM chips, it is critical to develop fully-secure, low-overhead, and scalable mitigation techniques. As such, 2020 and later years saw a surge in new RowHammer solutions. We briefly cover major recent ideas and directions in RowHammer mitigation.

Graphene~\cite{park2020graphene} uses the Misra-Gries algorithm~\cite{misra1982finding} for online frequent item counting~\cite{boyer1982fast,fischer1982finding,karp2003simple, fang1998computing} to track and identify frequently activated rows. Graphene then refreshes the rows neighboring those with activation counts that are close to the RowHammer threshold. This work, while effective, secure, and low-performance-overhead, requires large area overhead when scaled to address the worsening RowHammer vulnerability, as it uses area-expensive content addressable memory for storing metadata~\cite{yaglikci2021blockhammer}.
% , especially content-addressable memories, which can be costly.

\ignore{
%%% We should look at the literature here in future. Maybe we can do better than Misra-Gries?
~\omcomment{Also cite Karp's paper cited in our original ISCA 2014 paper. Also: no older references to online frequent item counting? Karp references a paper from 1968, but not sure if it is the same problem.}\atbcomment{we now cite karp, i could not find a citation to 1968 in karp}\agycomment{I added a work from Fang et al.~\cite{fang1998computing} that seems to be relevant and cited by Karp et al.~\cite{karp2003simple}. This one is published in 1998.}\omcomment{nothing from 1960s there?}\agycomment{I cannot see any. I've also checked the relevant papers that these two cite. Ataberk, could you also take a look?}\atbcomment{I already did. I could not find anything dating back to 1960s (looked at what misra gries and our two references from '82 cites.)}
}

BlockHammer~\cite{yaglikci2021blockhammer, blockhammergithub} uses counting Bloom Filters to identify frequently-activated rows and throttles accesses to those rows whose activation counts are close to the RowHammer threshold. BlockHammer thus does not require any proprietary information about DRAM chips (e.g., physical row adjacency information) and therefore can be implemented completely and securely in the memory controller. This work experimentally demonstrates that BlockHammer's performance overheads are low (similar to other best prior mechanisms) when there is no RowHammer attack. When there is a RowHammer attack, BlockHammer improves both system performance and energy consumption by throttling the attacker thread. This work also introduces a metric called RowHammer Likelihood Index (RHLI), and shows that this metric can be accurately computed to identify RowHammer attacks and report them to the system software. As such, BlockHammer provides an example of a system-level approach to mitigate RowHammer attacks' impact.\footnote{BlockHammer is freely and openly available~\cite{blockhammergithub}, along with six other RowHammer mitigation mechanisms~\cite{kim2014flipping, son2017making, you2019mrloc, lee2019twice, seyedzadeh2018cbt, park2020graphene}, implemented in Ramulator~\cite{ramulator,ramulatorgithub}.} This work deeply analyzes 14 major RowHammer defense mechanisms, with many quantitative and qualitative comparisons, and forms a basis for methodically comparing different RowHammer defenses.\footnote{The BlockHammer paper~\cite{yaglikci2021blockhammer} identifies four major desirable properties of a RowHammer defense, and demonstrates that BlockHammer is the only mechanism that satisfies all four properties. We refer the reader to Section~{9} in~\cite{yaglikci2021blockhammer} for more detail.}

\ignore{
\omcomment{Do we cite Panopticon, SofTRR and other mitigations, many of which we reviewed and discussed? DRAMSec papers are perfectly valid to cite. Let's not miss any works, as much as we can, regardless of how good or feasible the work is. This comment applies to exploiting and understanding thrusts as well. Please double check. Please also double check we cite all RH papers by major players, Microsoft (Stefan), Kaveh, Moin, ... Please give another pass quickly as we are very close to finalization.}\atbcomment{we now cite panopticon and moin's ecc paper (from 2021). we double checked.} \atbcomment{we now cite LTPIM.}
}

\ignore{
\agy{BlockHammer's GitHub repository~\cite{blockhammergithub} contains both register-transfer-level (RTL) implementation of BlockHammer's area-dominant components and simulator implementations of BlockHammer's and six other state-of-the-art RowHammer mitigation mechanisms~\cite{kim2014flipping, son2017making, you2019mrloc, lee2019twice, seyedzadeh2018cbt, park2020graphene}. BlockHammer also classifies 14 prior RowHammer mitigation mechanisms into four high-level approaches and qualitatively evaluates each mechanism in terms of four desirable properties of a RowHammer mitigation mechanism: 1)~comprehensive protection, 2)~compatibility with commodity DRAM chips, 3)~scalability with worsening RowHammer vulnerability, and 4)~deterministic protection. This qualitative evaluation shows that BlockHammer is the \emph{only} mechanism that posesses all four properties.}
}

More recently, Self-Managing DRAM (SMD)~\cite{hassan2022case, hassan2022smdgithub} takes a very different approach to RowHammer mitigation: SMD modifies the DRAM interface such that the DRAM chip can reject an activation command issued by the memory controller and thus gain time to perform internal maintenance operations, such as RowHammer mitigation. SMD observes that RowHammer mitigation is essentially a DRAM maintenance operation that can be best implemented within the DRAM chip based on device-level information available to DRAM manufacturers. SMD essentially provides ``breathing room'' (i.e., extra time) to the DRAM chip to autonomously implement such maintenance operations completely and securely within the DRAM chip, at low overheads. The results are promising: by carefully scheduling RowHammer mitigation actions at a fine enough granularity across different regions of a DRAM chip, in-DRAM RowHammer mitigation mechanisms inspired by PARA~\cite{kim2014flipping} and BlockHammer~\cite{yaglikci2021blockhammer} (i.e., SMD-PRP and SMD-PRP+) can lead to low performance and energy overheads~\cite{hassan2022case}.

Randomized Row Swap (RRS)~\cite{saileshwar2022randomized} and AQUA~\cite{saxena2022aqua} propose to relocate aggressor rows whose activation counts are close to the RowHammer threshold. These works adopt Graphene's approach to frequently-activated row detection. When such an aggressor row is detected, RRS swaps it with another randomly chosen row in the same bank, while AQUA moves the aggressor row into a dedicated quarantine region that stores the most frequently accessed rows. By doing so, both RRS and AQUA prevent a DRAM row from being activated enough times to induce a RowHammer bitflip. RRS and AQUA do \emph{not} need to know the physical layout of DRAM rows or make modifications to DRAM chips and thus they are compatible with commodity DRAM chips. On the downside, RRS and AQUA both perform data relocation over the memory bus, which requires off-chip data movement and exacerbates the already pressing data movement overhead in modern systems~\cite{mutlu2013memory, mutlu2020modern, boroumand2018google, deoliviera2021damov, boroumand2021google, kanev_isca2015, seshadri2017ambit}. Row-relocation based RowHammer defenses can be accelerated with in-DRAM data copy support~\cite{wang2020figaro,seshadri2013rowclone,hassan2019crow,chang2016low,hossein2020network, olgun2021pidram, olgun2021pidramgithub}.

Two recent works in 2022 aim to broadly reduce the overheads of RowHammer mitigations. HiRA~\cite{yaglikci2022hira} shows that the performance overheads of refresh-based in-DRAM RowHammer mitigation mechanisms can be reduced by performing refreshes in parallel with accesses and other refreshes~\cite{kim2012case,chang2014improving}, and this can be done, to an extent, in real DRAM chips. Hydra~\cite{qureshi2022hydra} shows that the hardware cost of counters used to track row activations can be reduced by storing such counters in DRAM and caching them in small structures on chip. These two works highlight the various overheads of RowHammer mitigation and take a step in improving efficiency of broad classes of mitigation techniques.

%)~inherit the hardware complexity issue of using CAM arrays for tracking %as in Graphene~\cite{park2020graphene} and 2)

\ignore{
~\cite{mutlu2013memory, mutlu2015research, dean2013tail, kanev_isca2015, ferdman2012clearing, wang2014bigdatabench, mutlu2019enabling, mutlu2019processing, mutlu2020intelligent, ghose.ibmjrd19, mutlu2020modern, boroumand2018google, wang2016reducing, pandiyan2014quantifying, koppula2019eden, kang2014co, mckee2004reflections, wilkes2001memory, kim2012case, wulf1995hitting, ghose.sigmetrics20, ahn2015scalable, PEI, hsieh2016transparent, wang2020figaro, deoliviera2021damov}.\agycomment{I copied these citations from DAMOV~\cite{deoliviera2021damov}.}
}

Many other works from 2020-2023 (e.g.,~\cite{yaglikci2021security, devaux2021method, frigo2020trrespass, kang2020cattwo, hassan2021utrr, qureshi2022hydra, kim2022mithril, lee2021cryoguard, marazzi2022protrr, zhang2022softtrr, joardar2022learning, juffinger2023csi, yaglikci2022hira, enomoto2022efficient, manzhosov2022revisiting, ajorpaz2022evax, naseredini2022alarm, joardar2022machine, zhang2020leveraging, loughlin2021stop, fakhrzadehgan2022safeguard, saroiu2022price, loughlin2022moesiprime, park2022rowhammer, han2021surround, bennett2021panopticon, qureshi2021rethinking, zhou2022ltpim, saroiu2022configure}) propose various other RowHammer mitigation mechanisms.  We refer the reader to the individual papers for more detail. As RowHammer is greatly worsening with DRAM technology scaling, we strongly believe that there is still a critical need and large potential to solve RowHammer at very low cost and very high efficiency, as we discuss in Section~\ref{sec:future_solutions}.

%~\cite{devaux2021method, yaglikci2021security}, \atb{ProTRR~\cite{marazzi2022protrr}}\omcomment{Can we cite more mechanisms here? What about BlockHammer}\agycomment{BlockHammer paper does not propose an in-DRAM solution. Maybe we can cite SMD~\cite{hassan2022case}?}

\ignore{
\omcomment{Giray, please describe row swapping approaches and more recent works from Moin -- please also check the Graphene and BlockHammer descriptions above and make better suggest changes. Please also add any other major works and conclude with a bulk citation to other works.} \agycomment{I added RRS and AQUA. I don't see a nice fit for Hydra~\cite{qureshi2022hydra}}

\omcomment{Giray: in addition to summarizing these works, it would be good to point out a few more things:\\
- BlockHammer's open source contribution \agycomment{done above}\\
- BlockHammer's catgorization of RH defenses and desired properties \agycomment{done above}\\
- What do we want from a good solution\atbcomment{Can yank from proposal}} \agycomment{Shouldn't this question be answered in Section 5.2?}
}

\ignore{
\omcomment{I copied these from an earlier section. They should all be cited here. And likely more should be cited.}
New solutions~\mitigatingRowHammerNewCitations{}. \agycomment{added a paragraph citing all of these.}
% \cite{park2020graphene, yaglikci2021security, yaglikci2021blockhammer, frigo2020trrespass, kang2020cattwo, hassan2021utrr, qureshi2022hydra, saileshwar2022randomized, kim2022mithril, lee2021cryoguard, marazzi2022protrr, zhang2022softtrr, joardar2022learning, juffinger2023csi, yaglikci2022hira, saxena2022aqua, enomoto2022efficient, manzhosov2022revisiting, ajorpaz2022evax, naseredini2022alarm, joardar2022machine, hassan2022case, zhang2020leveraging, loughlin2021stop, fakhrzadehgan2022safeguard, saroiu2022price, loughlin2022moesiprime}.
}

% \agycomment{I updated the mitigatingRowHammer macro below with the citations from TCAD paper.}
%%% CHECK-AG: Ditto here, but for mitigations.

% Mitigating RowHammer all works~\mitigatingRowHammerAllCitations{}.

% Mitigating RowHammer post-2020 works~\mitigatingRowHammerNewCitations{}.

\vspace{-0.5em}
\section{Future Directions} 
\label{sec:future}

Aside from continuing the path of prior research, including discovering and developing new RowHammer attacks and access patterns, we believe there are two major directions that are critical for future research to investigate and amplify efforts in.

\vspace{-0.5em}
\subsection{Building a Fundamental \& Comprehensive Understanding of RowHammer}

Even though there are various detailed characterization studies performed to understand various properties of RowHammer~\cite{kim2014flipping, park2016experiments, kim2020revisiting, orosa2021deeper, orosa2022spyhammer, cohen2022hammerscope, yaglikci2022understanding},
there is still a lot we do not know about RowHammer, its properties/sensitivities, and the manifestations of such properties in cutting-edge and future DRAM chips. It is critical to fundamentally understand the various properties of RowHammer under different conditions and access patterns, in order to develop fully-secure and efficient solutions (as we argue for in Section~\ref{sec:future_solutions}). 

\ignore{\omcomment{cite our and others' characterization studies}\agycomment{done}\omcomment{Does it make sense to cite spyhammer and hammerscope here?}\atbcomment{my thinking is that spyhammer is synergistic with deeperlook and hammerscope is with understanding reduced voltage in the understanding sense, so i would say yes}\agycomment{spyhammer already claims fine-grained rigorous characterization as one of its contributions, so I think it makes sense. The same applies to HammerScope too even though the paper's methodology is terrible and the data should be full of noise}}
\ignore{
\atbcomment{I think these are the three main properties we discuss in src proposal, but I had to expand on what we had a lot (writing is likely bad).}
\agycomment{reordering the three properties can help. First two are environmental conditions and access patterns --> which are studied but need more detailed analyses. The last one is aging, which is not studied yet.} \atbcomment{done. I ordered it the other way around 1) aging (nothing yet), 2) env. conds 3) access patterns (more work needed)}\omcomment{It is looking fine for now.}
}

We believe that at least several properties of RowHammer are critical to fundamentally understand going forward: sensitivity to 1) aging of DRAM chips, 2) environmental conditions (e.g., operating temperature, supply voltage), and 3) memory access patterns. There is no detailed characterization study that evaluates if and how the aging of a DRAM chip affects its RowHammer vulnerability. While some environmental conditions and memory access patterns are experimentally demonstrated to significantly affect the RowHammer vulnerability of a DRAM chip~\cite{kim2020revisiting,orosa2021deeper,yaglikci2022understanding,kogler2022half}, further research is needed to develop a more detailed understanding on the relationship between such properties and the RowHammer vulnerability of a DRAM chip. This understanding is necessary and critical to develop given that all existing RowHammer defense mechanisms rely on and future RowHammer defense mechanisms will likely rely on the measured RowHammer vulnerability of DRAM chips (e.g., the RowHammer threshold value) to securely prevent bitflips. Understanding the individual and combined effects of RowHammer's sensitivities to aging of DRAM chips, environmental conditions, and memory access patterns could yield accurate methodologies and infrastructures that can efficiently evaluate the RowHammer vulnerability of a given DRAM chip and facilitate the development of holistic solutions that completely prevent RowHammer across the entire computing system. We believe FPGA-based infrastructures for testing DRAM chips, such as SoftMC~\cite{hassan2017softmc,softmcgithub} and DRAM Bender~\cite{olgun2022drambender,safari-drambender}, are critical to enabling such studies, as they have been in the past.
%%% Can be removed for space, if needed: , as they have been in the past.

%\atbcomment{Do we want to plug in "correlation with other failure mechanisms" here? Copy paste from SRC: How does aging of DRAM/NVM chips affect the RH vulnerability? How does RH correlate with other types of failure mechanisms that are exacerbated due to aging?}\omcomment{No. We are out of space and we do not want to give too many hints for future research to others. Save this for a future extended version. Can you start a GDoc? We want to be more open in that GDoc as that will be for our research planning and brainstorming.}

\ignore{
%Thus, to efficiently and accurately determine the RowHammer threshold of a DRAM chip, a comprehensive understanding of RowHammer's sensitivity to environmental conditions, aging of DRAM chips, and access patterns is necessary. All existing RowHammer defense mechanisms rely on and future RowHammer defense mechanisms will likely rely on the RowHammer threshold value to securely prevent RowHammer bitflips. }

{and 4) correlation with other types of failure mechanisms (e.g., retention failures~\cite{dram-isca2013, kang-memforum2014, mutlu-imw13}).
To prevent RowHammer bitflips in a DRAM chip, . Accurately determining the RowHammer threshold of a DRAM chip requires extensive RowHammer characterization that takes environmental conditions, aging, and access patterns into account. Understanding the individual and combined effects of these properties could yield accurate methodologies and infrastructures that can efficiently evaluate the RowHammer vulnerability of a given system and facilitate development of holistic solutions that completely prevent RowHammer across the whole computing system. } \atbcomment{WIP}...\omcomment{ataberk, please continue this... include the effect of aging on rowhammer... and many other things we have in the proposal. A lot from the proposal can be leveraged, but do it well.}}

We also believe that it is important to more profoundly understand the effects of RowHammer on real systems and real applications, both malicious and non-malicious. To this end, it is critical to do research that pushes the boundaries of generating RowHammer bitflips on many different types of systems, including mobile and server CPUs, GPUs, accelerators, FPGAs, as well as different DRAM and memory types, including HBM and emerging NVM technologies. Such research can not only discover new problems and sensitivities but also pave the way to generalized solutions that are applicable to many systems.  

\vspace{-0.5em}
\subsection{Designing Extremely Efficient Solutions to RowHammer}
\label{sec:future_solutions}
As the RowHammer vulnerability worsens with DRAM technology scaling, developing extremely efficient and fully-secure RowHammer solutions becomes increasingly important. Even though many prior works develop various software- and hardware-level RowHammer solutions, these solutions incur non-negligible and increasingly more significant system performance, energy, and hardware area overheads as RowHammer vulnerability worsens.

We believe that developing new low-cost (in terms of performance, energy, area) and provably-secure RowHammer solutions is critical to efficiently preventing RowHammer bitflips going forward. As DRAM continues to scale, RowHammer bitflips can occur at smaller activation counts and thus a benign workload's DRAM row activation rates can approach or even exceed the RowHammer threshold~\cite{loughlin2022moesiprime,saxena2022aqua, saileshwar2022randomized, qureshi2022hydra, yaglikci2021blockhammer}. Thus, a system may experience bitflips or frequently trigger RowHammer defense mechanisms even without a malicious party performing a RowHammer attack in the system, leading to data corruption or significant performance degradation. To avoid such problems, we advocate co-architecting of the system and the memory together. A holistic solution that takes a system-memory co-design approach (as advocated earlier by~\cite{mutlu2013memory,kim2014flipping,patel2022case,mutlu2015research,kang2014co}) can both prevent RowHammer bitflips and detect RowHammer attacks while at the same time avoiding potential performance and denial-of-service problems due to both RowHammer attacks and RowHammer mitigation mechanisms. For example, such a system can efficiently relocate/isolate data or throttle/relocate/isolate threads such that the performance of non-malicious applications is unaffected by RowHammer attacks or mitigations.\footnote{BlockHammer~\cite{yaglikci2021blockhammer, blockhammergithub} takes a step towards this direction by throttling the threads that are identified as RowHammer attacks. We believe that there is more to be done to further reduce the performance problems due to both Rowhammer attacks and RowHammer mitigations.} With worsening RowHammer vulnerability, such holistic solutions could pave the way to extremely efficient and fully-secure defenses against RowHammer.

\ignore{
\omcomment{I revised the above paragraphs. Please check. Also, do we want to give out these - I guess it is not bad but please think.}\atbcomment{looks good, i'm not sure either about giving this out.}\agycomment{These paragraphs look good to me too. I think the word ``relocating'' makes the research direction a bit too specific in case you have doubts about giving the idea away. Instead, we can say something like: ``reduce the interference between the memory accesses of benign workloads and identified RowHammer attack''... It is a broader direction which includes both relocating them and BlockHammer's approach of throttling threads that perform RowHammer attack.}\omcomment{take a cut}\agycomment{I drafted an alternative text above and marked the previous version that can be replaced by the alternative.}
}

% free the resources that would have been wasted by a malicious RowHammer attack for concurrently running benign applications, similar to what BlockHammer~\cite{blockhammergithub, yaglikci2021blockhammer} does.
% to the RowHammer problem can not only prevent RowHammer bitflips but also 
%  and

% \agy{For example, BlockHammer~\cite{yaglikci2021blockhammer, blockhammergithub} partially achieves this by freeing the memory bandwidth that otherwise would have been wasted by a RowHammer attack for concurrently running benign applications. However, we believe that there is a large headroom for improvement.}

% \atbcomment{I tried to zip Giray's paragraph, but it could go better in a more general message. I feel it is too specialized to start a future work paragraph.}

We also believe that more flexible and efficient RowHammer solutions can take advantage of the wide variation in RowHammer vulnerability (as shown by~\cite{kim2020revisiting, orosa2021deeper, hassan2021utrr, yaglikci2022understanding}) across 1)~cells in a DRAM chip, 2)~DRAM chips, 3)~manufacturers, 4)~DRAM types and generations, 5) environmental conditions, and 6) data patterns, to statically and dynamically adapt to system and workload characteristics. In current practice, RowHammer solutions need to be configured for the DRAM chip with the smallest RowHammer threshold. Such solutions are overly aggressive, as they try to cover the worst possible case, i.e., the most RowHammer-vulnerable DRAM chip that is acceptable to be sold, and can therefore induce large system performance and energy overheads that are unnecessary in the common case. We believe future RowHammer solutions need to be easily (re)configurable or programmable, such that they can statically and dynamically adapt to system and workload characteristics, and there is significant research needed towards enabling such flexible and efficient solutions.\footnote{We believe the PARA defense mechanism~\cite{kim2014flipping} could be a suitable starting point for exploring such (re)configurable yet low-cost solutions due to its 1)~fixed hardware cost that does {\em not} increase with worsening RowHammer vulnerability and 2)~easily reconfigurable nature where updating the probability threshold directly tunes its aggressiveness, as described in our recent HiRA work at MICRO 2022~\cite{yaglikci2022hira}.}

\ignore{
\omcomment{I revised the above. Please check. Do we want to give out this direction? We should be the first one writing a strong paper in this, but I am fine with giving it out if you are confident we will be done soon. We should be.} \atbcomment{I fixed one numbering issue. I am confident that we will very soon write strong papers on this.}\agycomment{I don't think that this is giving away an idea. It merely restates a desirable RowHammer defense property, which we already state in HiRA (first paragraph of Section 9).}
}
\ignore{
\omcomment{Ataberk, please take a cut at this section, too. We need to start it nicely. Please write and I will look later.}\atbcomment{done}\agycomment{done}
}

\ignore{
%%% JEREMIE-DONE: Double check LPDDR4 below...
We believe there is a lot more research to come that will build on
RowHammer, from at least three perspectives: 1) the security attack
perspective, 2) the defense/mitigation perspective, 3) a broader
understanding, modeling, and prevention \jk{perspective}.

As systems security researchers understand more about RowHammer, and
as the RowHammer phenomenon continues to fundamentally affect memory
chips due to technology scaling problems~\cite{onur-date17},
researchers and practitioners will develop different types of attacks
to exploit RowHammer in various contexts and in many more creative
ways. RowHammer is a critical problem that manifests in the
difficulties in DRAM scaling and is expected to only become worse in
the future~\cite{mutlu13, mutlu2015main, superfri14}. As we discussed,
some recent reports suggest that new-generation DRAM chips are
vulnerable to RowHammer (e.g.,
DDR4~\cite{rowhammer-thirdio,pessl2016drama, aga2017good,
  aichinger2015ddr},
ECC~\cite{rowhammer-isca2014,cojocar19exploiting},
LPDDR3 and LPDDR2~\cite{drammer}). This indicates that effectively mitigating the
RowHammer problem with low overhead is difficult and becomes more
difficult as process technology scales further. Even with the wide
array of works that build on top of RowHammer, we believe that these
papers have yet to scratch the surface of this field of reliability
and security, especially as manufacturing technology scaling continues
in all technologies. It is critical to deeply understand the
underlying factors of the RowHammer problem (and more generally the
crosstalk problem) such that we can effectively prevent these issues
across all technologies with minimal overhead.  As DRAM cells become
even smaller and less reliable, it is likely for them to become even
more vulnerable to complicated and different modes of failure that
are sensitized only under specific access-patterns and/or
data-patterns. As a scalable solution for the future, our ISCA 2014
paper argues for adopting a system-level approach~\cite{mutlu-imw13}
to DRAM reliability and security, in which the DRAM chips, the memory
controller, and perhaps the operating system collaborate together to
diagnose/treat emerging DRAM failure modes.

We believe that more and more researchers will focus on providing
security in all aspects of computing so that such hardware faults that are exposed to the software (and thus the public) are minimized. RowHammer enabled a shift of mindset among mainstream security researchers: general-purpose
hardware is fallible (in a very widespread manner) and its problems
are actually exploitable. This shift of mindset enabled many systems
security researchers to examine hardware in more depth and understand
its inner workings and vulnerabilities better. We believe it is no
coincidence that two of the groups that concurrently discovered the
heavily-publicized Meltdown~\cite{lipp2018meltdown} and
Spectre~\cite{kocher2018spectre} vulnerabilities (Google Project Zero
and TU Graz InfoSec) have heavily worked on RowHammer attacks
before. We believe this shift in mindset, enabled in good part by the
existence and prevalence of RowHammer, will continue to be very be
important for discovering and solving other potential vulnerabilities
that may rise as a result of both technology scaling and hardware
design.

\subsection{Other Potential Vulnerabilities}
\label{sec:other-problems}
\label{sec:future-other-problems}

%%% JEREMIE-DONE: Please expand the references in the below part to be more inclusive of our newer works, especially, of course, as appropriate to the phrase where they are referenced.
We believe that, as memory technologies scale to higher densities,
other problems may start appearing (or may already be going unnoticed)
that can potentially threaten the foundations of secure systems. There
have been recent large-scale field studies of memory errors showing
that both DRAM and NAND flash memory technologies are becoming less
reliable~\cite{superfri14,
  justin-memerrors-dsn15,dram-field-analysis2, dram-field-analysis3,
  dram-field-analysis4, justin-flash-sigmetrics15,
  flash-field-analysis2, cai2017errors, cai2017error,
  luo2018improving, luo2018heatwatch, cai-date12, cai-hpca15,
  mutlu-imw13, patel2017reach, onur-date17}. As detailed experimental
analyses of real DRAM and NAND flash chips show, both technologies are
becoming much more vulnerable to cell-to-cell interference
effects~\cite{superfri14, rowhammer-isca2014, cai-dsn15,
  cai-sigmetrics14, cai-iccd13, cai-date12,cai-date13, flash-fms-talk,
  yixin-jsac16, cai-hpca17, cai2017errors, cai2017error, mutlu-imw13,
  onur-date17}, data retention is becoming significantly more
difficult in both technologies~\cite{raidr,samira-sigmetrics14,
  dram-isca2013, khan-dsn16, avatar-dsn15, darp-hpca2014,
  kang-memforum2014, mandelman-jrd02, cai-hpca15, cai-iccd12,
  warm-msst15, cai-date12,cai-date13,cai-itj2013, flash-fms-talk,
  memcon-cal16, cai2017errors, cai2017error, luo2018improving,
  luo2018heatwatch, superfri14, mutlu-imw13}, and error variation
within and across chip, and across operating conditions, is increasingly prominent~\cite{dram-isca2013,
  aldram, kevinchang-sigmetrics16, dram-process-variation-3,
  cai-date12, cai-date13, lee2017design, kim2018solar, kim2018dram,
  kim2019drange}.  Emerging memory technologies~\cite{mutlu-imw13,
  meza-weed13}, such as Phase-Change Memory~\cite{pcm-isca09,
  zhou-isca09, moin-isca09, moin-micro09, wong-pcm, raoux-pcm,
  pcm-ieeemicro10, pcm-cacm10, justin-taco14, rbla},
STT-MRAM~\cite{chen-ieeetmag10,kultursay-ispass13}, and
RRAM/ReRAM/memristors~\cite{wong-rram} are likely to exhibit similar
and perhaps even more exacerbated reliability issues. We believe, if
not carefully accounted for and corrected, these reliability problems
may surface as security problems as well, as in the case of RowHammer,
especially if the technology is employed as part of the main memory
system that is directly exposed to user-level programs.

We briefly examine two example potential vulnerabilities. We believe
future work examining these vulnerabilities, among others, are
promising for both fixing the vulnerabilities and enabling the
effective scaling of memory technology.

\subsubsection{Data Retention Failures}

Data retention is a fundamental reliability problem, and hence a
potential vulnerability, in especially charge-based memories like DRAM and flash memory. This is because charge leaks out of the charge storage unit
(e.g., the DRAM capacitor or the NAND flash floating gate) over
time. As such memories become denser, three major trends make data
retention more difficult~\cite{raidr,dram-isca2013,kang-memforum2014,
cai-hpca15}. First, the number of memory cells increases, leading to
the need for more refresh operations to maintain data
correctly. Second, the charge storage unit (e.g., the DRAM capacitor)
becomes smaller and/or morphs in structure, leading to potentially
lower retention times. Third, the voltage margins that separate one
data value from another become smaller (e.g., the same voltage window
gets divided into more ``states'' in NAND flash memory, to store more
bits per cell), and, as a result, the same amount of charge loss is more
likely to cause a bit error in a smaller technology node than in a larger
one.

\textbf{DRAM Data Retention Issues}

Data retention issues in DRAM are a fundamental scaling limiter of the
DRAM technology~\cite{mutlu-imw13, dram-isca2013, kang-memforum2014}.  We have
shown, in recent works based on rigorous experimental analyses of
modern DRAM chips~\cite{dram-isca2013, samira-sigmetrics14,
  avatar-dsn15, khan-dsn16, memcon-micro17, patel2017reach}, that
determining the minimum retention time of a DRAM cell is getting
significantly more difficult. Thus, determining the correct rate at
which to refresh DRAM cells has become more difficult, as also
indicated by industry~\cite{kang-memforum2014}. This is due to two
major phenomena, both of which get worse (i.e., become more prominent)
with technology scaling. First, Data Pattern Dependence (DPD): the
retention time of a DRAM cell is heavily dependent on the data pattern
stored in itself and in the neighboring
cells~\cite{dram-isca2013}. Second, Variable Retention Time (VRT): the
retention time of some DRAM cells can change drastically over time,
due to a memoryless random process that results in very fast charge
loss via a phenomenon called trap-assisted gate-induced drain
leakage~\cite{yaney1987meta, restle1992dram, dram-isca2013}. These
phenomena greatly complicate the accurate determination of minimum
data retention time of DRAM cells. In fact, VRT, as far as we know, is
very difficult to test for because there seems to be no way of
determining that a cell exhibits VRT until that cell is observed to
exhibit VRT and the time scale of a cell exhibiting VRT does not seem
to be bounded, given the current experimental
data~\cite{dram-isca2013,samira-sigmetrics14,avatar-dsn15,patel2017reach}. As
a result, some retention errors can easily slip into the field because
of the difficulty of the retention time testing.  Therefore, data
retention in DRAM is a vulnerability that can greatly affect both
reliability and security of current and future DRAM generations. We
encourage future work to investigate this area further, from both
reliability and security, {\em as well as} performance and energy
efficiency perspectives.  Various works in this area provide insights
about the retention time properties of modern DRAM devices based on
experimental data~\cite{dram-isca2013, samira-sigmetrics14,
  avatar-dsn15, khan-dsn16, memcon-micro17, softmc, patel2017reach},
develop infrastructures to obtain valuable experimental
data~\cite{softmc}, and provide potential solutions to the DRAM
retention time problem~\cite{raidr, dram-isca2013,
  samira-sigmetrics14, avatar-dsn15, khan-dsn16, memcon-cal16,
  memcon-micro17, darp-hpca2014, patel2017reach}, all of which the
future works can build on.

Note that data retention failures in DRAM are likely to be
investigated heavily to ensure good performance and energy efficiency.
And, in fact they already are being investigated for this purpose
(see, for example,~\cite{raidr, darp-hpca2014, memcon-cal16,
  samira-sigmetrics14, avatar-dsn15, khan-dsn16, memcon-micro17, patel2017reach}).
We believe it is important for such investigations to ensure no new
vulnerabilities (e.g., side channels) open up due to the solutions
developed.

% DPD
% VRT
% Basically much more difficult
% This vulnerability can lead to security vulnerabilities.
% Needs to be fixed anyway for perf/scaling reasons, but when fixing we should not open up new vuknetrabikities

\textbf{NAND Flash Data Retention Issues}

Experimental analysis of modern flash memory devices show that the
dominant source of errors in flash memory are data retention
errors~\cite{cai-date12,cai2017error}. As a flash cell wears out, its
charge retention capability degrades~\cite{cai-date12, cai-hpca15,
  cai2017errors, cai2017error, luo2018improving, luo2018heatwatch,
  justin-flash-sigmetrics15, flash-field-analysis2} and the cell
becomes leakier.  As a result, to maintain the original data stored in
the cell, the cell needs to be refreshed~\cite{cai-iccd12,
  cai-itj2013}. The frequency of refresh increases as wearout of the
cell increases. We have shown that performing refresh in an adaptive
manner greatly improves the lifetime of modern MLC (multi-level cell)
NAND flash memory while causing little energy and performance
overheads~\cite{cai-iccd12, cai-itj2013}. Most high-end SSDs today
employ such adaptive refresh mechanisms.

As flash memory scales to smaller manufacturing technology nodes and
even more bits per cell, data retention becomes a bigger problem. As
such, it is critical to understand the issues with data retention in
flash memory. Our recent work provides detailed experimental analysis
of data retention behavior of planar and 3D MLC NAND flash
memory~\cite{cai-hpca15, cai2017errors, cai2017error,
  luo2018improving, luo2018heatwatch}. We show, among other things,
that there is a wide variation in the leakiness of different flash
cells: some cells leak very fast, some cells leak very slowly. This
variation leads to new opportunities for correctly recovering data
from a flash device that has experienced an uncorrectable error: by
identifying which cells are fast-leaking and which cells are
slow-leaking, one can probabilistically estimate the original values
of the cells before the uncorrectable error occurred. This mechanism,
called {\em Retention Failure Recovery}, leads to significant
reductions in bit error rate in modern MLC NAND flash
memory~\cite{cai-hpca15, cai2017errors, cai2017error} and is thus very
promising. Unfortunately, it also \jk{points to} a potential security
and privacy vulnerability: by analyzing data and cell properties of a
failed device, one can potentially recover the original data. We
believe such vulnerabilities can become more common in the future and
therefore they need to be anticipated, investigated, and understood.

\subsubsection{Other Vulnerabilities in NAND Flash Memory}

We believe other sources of error (e.g., cell-to-cell interference) and
cell-to-cell variation in flash memory can also lead various vulnerabilities.
For example, another type of variation (that is similar to the variation in
cell leakiness that we described above) exists in the vulnerability of flash
memory cells to read disturbance~\cite{cai-dsn15}: some cells are much more
prone to read disturb effects than others. This wide variation among cells
enables one to probabilistically estimate the original values of cells in flash
memory after an uncorrectable error has occurred. Similarly, one can
probabilistically correct the values of cells in a page by knowing the values
of cells in the neighboring page~\cite{cai-sigmetrics14}. These
mechanisms~\cite{cai-dsn15, cai-sigmetrics14} are devised to improve flash
memory reliability and lifetime, but the same phenomena that make them
effective in doing so can also lead to potential vulnerabilities, which we
believe are worthy of investigation to ensure security and privacy of data in
flash memories.

As an example, we have recently shown~\cite{cai-hpca17} that it is
theoretically possible to exploit vulnerabilities in flash memory
programming operations on existing solid-state drives (SSDs) to cause
(malicious) data corruption. This particular vulnerability is caused
by the {\em two-step programming} method employed in dense flash
memory devices, e.g., MLC NAND flash memory. An MLC device partitions
the threshold voltage range of a flash cell into four
distributions. In order to reduce the number of errors introduced
during programming of a cell, flash manufacturers adopt a two-step
programming method, where the least significant bit of the cell is
partially programmed first to some intermediate threshold voltage, and
the most significant bit is programmed later to bring the cell up to
its full threshold voltage.  We find that two-step programming exposes
new vulnerabilities, as both cell-to-cell program interference and
read disturbance can disrupt the intermediate value stored within a
multi-level cell before the second programming step completes. We show
that it is possible to exploit these vulnerabilities on existing
solid-state drives (SSDs) to alter the partially-programmed data,
causing (malicious) data corruption. We experimentally characterize
the extent of these vulnerabilities using contemporary 1X-nm (i.e.,
15-19nm) flash chips~\cite{cai-hpca17}. Building on our experimental
observations, we propose several new mechanisms for MLC NAND flash
that eliminate or mitigate disruptions to intermediate values,
removing or reducing the extent of the vulnerabilities, mitigating
potential exploits, and increasing flash lifetime by
16\%~\cite{cai-hpca17}. We believe investigation of such
vulnerabilities in flash memory will lead to more robust flash memory
devices in terms of both reliability and security, as well as
performance. In fact\jk{,} a recent work from IBM builds on our
work~\cite{cai-hpca17} to devise a security attack at the file system
level~\cite{ibm-fs-attack}.

\subsection{Prevention}
\label{sec:prevention}
\label{sec:future-prevention}

Various reliability problems experienced by scaled memory technologies, if not
carefully anticipated, accounted for, and corrected, may surface as security
problems as well, as in the case of RowHammer.  We believe it is critical to
develop principled methods to understand, anticipate, and prevent such
vulnerabilities. In particular, principled methods are required for three major
steps in the design process.

First, it is critical to understand the potential failure mechanisms and
anticipate them beforehand. To this end, developing solid methodologies for
failure modeling and prediction is critical. To develop such methodologies, it
is essential to have real experimental data from past and present devices. Data
available both at the small scale (i.e., data obtained via controlled testing
of individual devices, as in, e.g.,~\cite{dram-isca2013, aldram, samira-sigmetrics14, kevinchang-sigmetrics16, cai-date12, cai-date13, cai-dsn15, cai-hpca15, yixin-jsac16, cai2017errors, cai2017error, cai-iccd12, cai-itj2013, cai-iccd13, cai-sigmetrics14, cai-hpca17, yucai-thesis, luo2018improving, luo2018heatwatch, patel2019understanding, kim2018solar, patel2017reach, kim2019drange, kim2018dram}) as well as at the large scale (i.e., data
obtained during in-the-field operation of the devices, under
likely-uncontrolled conditions, as in, e.g.,~\cite{justin-memerrors-dsn15,
justin-flash-sigmetrics15}) can enable accurate models for failures, which
could aid many purposes, including the development of better reliability
mechanisms and prediction of problems before they occur.

Second, it is critical to develop principled architectural methods that can
avoid, tolerate, or prevent such failure mechanisms that can lead to
vulnerabilities. For this, we advocate co-architecting of the system and the
memory together, as we described earlier. Designing intelligent, flexible,
configurable, programmable, patch-able memory controllers that can understand and correct existing and
potential failure mechanisms can greatly alleviate the impact of failure
mechanisms on reliability, security, performance, and energy efficiency. A {\em
system-memory co-design} approach can also enable new opportunities, like
performing effective processing near or in the memory device
(e.g.,~\cite{rowclone, vivek-and-or, tesseract, pei, gs-dram, tom-isca16,
impica-iccd16, lazypim, pattnaik-pact16, emc-isca16, 
ghose2018enabling, boroumand2018google, kim2018grim, seshadri2017ambit,
liu2017concurrent, seshadri2017simple, aga2017compute, akin2015data,
asghari2016chameleon, babarinsa2015jafar, lisa, chi2016prime, farmahini2015nda,
gao2015practical, gao2016hrl, gu2016biscuit, guo20143d, cre-micro16,
hassan2015near, hsieh2016accelerating, sramsod, kim2016neurocube,
kim2017toward, lee2015bssync, li2017drisa, pinatubo, loh2013processing,
morad2015gp, nai2017graphpim, pugsley2014ndc, shafiee2016isaac,
sura2015data, zhang2014top, zhu2013accelerating, mutlu2019processing, singh2019napel, mutlu2019enabling, boroumand2019conda}). In addition to designing the
memory device together with the controller, we believe it is important to
investigate mechanisms for good partitioning of duties across the various
levels of transformation in computing, including system software, compilers,
and application software.
%%% JEREMIE-DONE: Please add to the list of processing-in-memory works above. It seems like it stopped in 2016. Please include all our works and works from others as well. Our book chapters should also be inluded (we have one driven by Saugata)
%%% JEREMIE-COMMENT: What is the book publishing Saugata's chapter? Right now I can only find it on arxiv. 

Third, it is critical to develop principled methods for electronic
design, automation and testing, which are in harmony with the failure
modeling/prediction and system reliability methods, which we mentioned
in the above two paragraphs. Design, automation and testing methods
need to provide high and predictable coverage of failures and work in
conjunction with architectural and across-stack mechanisms. For
example, enabling effective and low-cost {\em online profiling of
  DRAM}~\cite{dram-isca2013, samira-sigmetrics14, avatar-dsn15,
  khan-dsn16, memcon-cal16, lee2017design, patel2017reach} in a
principled manner requires cooperation of failure modeling mechanisms,
architectural methods, and design, automation and testing methods.
%%% JEREMIE-DONE: Donghyuk's DIVA-DRAM work in SIGMETRICS 2017 work does not seem to be cited. Can you include that above and in all appropriate places inthe document?  
%%% JEREMIE-DONE: Could you also check other works?

}
% \vspace{-0.5em}
\section{Conclusion}

We provided a brief overview of the history and current state of research and development on the RowHammer vulnerability. We described major future research directions which we believe are critical to fundamentally understanding and solving RowHammer. We conclude that even though much research and development is done on the topic, a lot more needs to be done going forward, since RowHammer is a fundamental DRAM technology scaling problem that is getting worse in newer DRAM chips that will continue to be employed across almost all computing systems. We hope the discussion and ideas provided in this paper provide a useful path for the community to find ways of fundamentally understanding and efficiently solving the RowHammer problem.

\ignore{
%%% ONUR-TODO: Revise the conclusion saying that ``we provided a
%%% retrospective on RowHammer and a survey of many flourishing works
%%% that build on it. ...'' I will need to revise this one...

%%% ONUR-TODO: Use the best ideas and storyline from my RowHammer
%%% talks, especially the last one at TopinHES

We provided a retrospective on the RowHammer problem and our original
ISCA 2014 paper~\cite{rowhammer-isca2014} that introduced the problem,
and a survey of many flourishing works that have built on
RowHammer. It is clear that the reliability of memory technologies we
greatly depend on is reducing, as these technologies continue to scale
to ever smaller technology nodes in pursuit of higher densities. These
reliability problems, if not anticipated and corrected, can also open
up serious security vulnerabilities, which can be very difficult to
defend against, if they are discovered in the field. RowHammer is an
example, likely the first one, of a hardware failure mechanism that
causes a practical and widespread system security vulnerability. As
such, its implications on system security research are tremendous and
exciting. We hope the summary, retrospective, and commentary we
provide in this paper on the RowHammer phenomenon are useful for
understanding the RowHammer problem, its context, mitigation
mechanisms, and the large body of work that has built on it in the
past five years.

We believe that the need to prevent such reliability and security
vulnerabilities at heavily-scaled memory technologies opens up new
avenues for principled approaches to 1) understanding, modeling, and
prediction of failures and vulnerabilities, and 2) architectural as
well as design, automation and testing methods for ensuring reliable
and secure operation. We believe the future is very bright for
research in reliable and secure memory systems, and many discoveries
abound in the exciting yet complex intersection of reliability and
security issues in such systems.

}

% \vspace{-0.5em}
\begin{acks}

We thank the organizers of the ASP-DAC 2023 conference for the invitation to contribute this invited paper and deliver an associated invited talk. We acknowledge many SAFARI Research Group Members who have contributed to some of the works described in this paper, especially Jeremie Kim and Hasan Hassan. We thank all members of the SAFARI Research Group for the stimulating and scholarly intellectual environment they provide. We acknowledge the generous gift funding provided by our industrial partners (especially by Google, Huawei, Intel, Microsoft, VMware), which has been instrumental in enabling the decade+ long research we have been conducting on RowHammer. This work was in part supported by the Microsoft Swiss Joint Research Center.

\ignore{
This paper is based on two previous papers we have written on RowHammer, one that first scientifically introduced and analyzed the phenomenon in ISCA 2014~\cite{rowhammer-isca2014} and the other that
provided an analysis and future outlook on RowHammer~\cite{onur-date17}. \jkthree{The presented work} is a result
of the research done together with many students and collaborators over the
course of the past eight years. In particular, three PhD theses have shaped the
understanding that led to this work. These are Yoongu Kim's thesis entitled
"Architectural Techniques to Enhance DRAM Scaling"~\cite{yoongu-thesis}, Yu Cai's thesis entitled "NAND Flash Memory: Characterization, Analysis, Modeling and Mechanisms"~\cite{yucai-thesis} and his continued follow-on work after his thesis, summarized in~\cite{cai2017error, cai2017errors}, and Donghyuk Lee's thesis entitled "Reducing DRAM Latency at Low Cost
by Exploiting Heterogeneity"~\cite{donghyuk-thesis-arxiv16}. We also acknowledge various funding agencies (NSF, SRC, ISTC, CyLab) and industrial partners (AliBaba, AMD, Google,
Facebook, HP Labs, Huawei, IBM, Intel, Microsoft, Nvidia, Oracle, Qualcomm,
Rambus, Samsung, Seagate, VMware) who have supported the presented and
other related work in \jkthree{our} group generously over the years.

The first version of
\jkthree{the talk associated with this paper} was delivered at a CMU CyLab Partners Conference in September
2015. \jkthree{Other versions of the talk were} delivered as part of an Invited Session
at DAC 2016, with a collaborative accompanying paper entitled ''Who Is the
Major Threat to Tomorrow\jkfour{'}s Security? You, the Hardware Designer''~\cite{dac-invited-paper16}, \jkthree{at DATE 2017~\cite{onur-date17},} and at the Top Picks in Hardware and
Embedded Security workshop, co-located with ICCAD 2018~\cite{TopPicks}, where RowHammer was selected as a Top Pick among hardware and embedded security papers
published between 2012-2017. \jkthree{The most recent version of the associated talk was delivered at COSADE 2019~\cite{mutlu2019rowhammer}}.
}

\end{acks}

\balance

\Urlmuskip=0mu plus 1mu\relax
% \setstretch{0.95}
% \bibliographystyle{ACM-Reference-Format}
\bibliographystyle{IEEETranS}
\bibliography{paper}

\end{document}